# Missing Baryons and the Warm-Hot Intergalactic Medium

Fabrizio Nicastro[1,2], Smita Mathur[3], Martin Elvis[1]

1. Harvard-Smithsonian Center for Astrophysics, Cambridge, MA, 02138, USA

2. Osservatorio Astronomico di Roma – Istituto Nazionale di Astrofisica, I-00040Rome, Italy

3. Ohio State University, Columbus, OH, 43210, USA

**Stars and gas in galaxies, hot intracluster medium, and intergalactic photo-ionized gas make up at most half of the baryons that are expected to be present in the universe. The majority of baryons are still missing and are expected to be hidden in a web of warm-hot intergalactic medium. This matter was shock-heated during the collapse of density perturbations that led to the formation of the relaxed structures that we see today. Finding the missing baryons and thereby producing a complete inventory of possibly the only detectable component of the energy-mass budget of the universe is crucial to validate or invalidate our standard cosmological model.**

Despite recent progress in cosmology in assaying the energy and mass budget of the universe, very little is still known about the nature and origin of most of its constituents. Within the framework of our standard cosmological model (SCM) (*1,2*), most (95%) of the universe (or $\Omega$, the mass density of the universe divided by the critical density required to close the Universe) is composed primarily of "dark" energy (70 %) and "dark" matter (25 %), both dubbed "dark" as a reflection of our inability to directly detect and identify them. Less well known is that the situation is only marginally better for the universe's remaining 5 % of detectable matter.

As baryons – the protons and atomic nuclei that constitute the ordinary matter of which stars, planets, and ourselves are made – this remaining matter should, in principle, be in a form that we can detect and measure in its physical state. We know from absorption line spectroscopy of distant quasars that clouds of baryons were present in the early universe about 10 billion years ago (redshift $z \simeq 2$) (*3,4*), in the form of photo-ionized diffuse intergalactic gas, accounting for at least three-quarters of the total baryon mass in the universe as inferred by both cosmic microwave background anisotropies (*1,2*) and "big-bang nucleosynthesis" predictions when combined with observed light-element ratios at $z > 2$ (*5*) ($\Omega_b > 3.5$ %, i.e. > 73 % of the estimated baryon mass in the Universe). However, these clouds of photo-ionized intergalactic gas have become more and more sparse as time moved towards the present and structures (galaxies, galaxy groups, and clusters) started to be assembled. Only a small fraction of the baryons that were present in the intergalactic medium at $z > 2$ are now found in stars, cold or warm interstellar matter, hot intracluster gas and residual photo-ionized intergalactic medium. Today we can account for less than 50 % of the baryon mass predicted by the SCM, implying that at least 50 % of the baryons are now "missing" (*6,7*) (Figure 1).

The leading theory of cosmological structure formation (known as Λ-CDM; or, cold dark matter models including dark energy, designated by the cosmological constant Λ) predicts that, as the universe evolves toward present and density perturbation grows to form structures, baryons in the diffuse intergalactic medium accelerate toward the sites of structure formation under the growing influence of gravity and go through shocks that heat them to temperatures of millions of kelvin. These "missing baryons" may have become difficult to detect by being concentrated in a filamentary web of tenuous (baryon density $n_b \simeq 10^{-6}$-$10^{-5}$ cm$^{-3}$, corresponding to overdensities of $\delta = n_b/\langle n_b \rangle \simeq$ 5-50, compared with the average baryon density in the universe $\langle n_b \rangle = 2 \times 10^{-7}$ cm$^{-3}$) warm-hot intergalactic medium (WHIM) that has been continuously shock-heated during the process of structure formation: [e.g. (*8*)]. This matter is hot ($10^5$-$10^7$ K) and is so highly ionized that it can only absorb or emit far-ultraviolet (FUV) and soft x-ray photons, primarily at lines of highly ionized (Li-like, He-like or H-like) C, O, Ne and Fe [e.g. (*9*)].

If identifying the mysterious dark energy and dark matter is challenging, then the problem of "missing baryons" is more acute. One could even argue that baryons represent the only directly detectable component of the predicted mass-energy budget of the universe [e.g. (*10, 11*)]. Moreover, determining the physical conditions, metal (heavy element) content, dynamics and kinematics of these baryons, as a function of cosmic time, gives us a unique set of tools to study the evolution of large scale structure (LSS) in the universe and the feedback mechanisms of gas ejected from galaxies and quasars onto LSS in the framework of such models [e.g. (*12*)].

This program of tracking the cosmic baryons can only be carried out by exploiting the synergies between multiwavelength observational approaches and theoretical computations: high resolution x-ray and UV spectroscopy measures (i) relative and absolute metal content , (ii) ionization correction, and so (iii) mass of the WHIM filaments; optical and infrared multi-filter photometry and spectroscopy measure the galaxy density around WHIM filaments, and so map dark-matter concentrations in the universe. These observations can then be used to feed and hence refine new simulations, to study and address the fundamental problem of feedback between virialized structures and the surrounding diffuse intergalactic medium.  All this has only relatively recently become possible thanks to the advent of sufficiently high-resolution x-ray and UV optics [*Chandra* and *Hubble Space Telescope (HST)*] and spectrometers [gratings on *Chandra*, HST, XMM-*Newton* p(XMM, X-ray Multimirror Mission) plus the Far Ultraviolet Spectroscopic Explorer (FUSE)], on one hand, and the rapidly increasing power of today's computational resources (such as high-resolution hydro-dynamical simulations, refined and extended atomic-data computations, non-equilibrium post-shock ionization modelling), on the other.

Due to the extreme low density (1-10 particles per cubic meter) and relatively small size (1-10 Mpc) of the WHIM filaments, the intensity of the signal expected from the WHIM observables (either in emission or in absorption) is low, both in the UV and in the x-ray band. This makes the search for the Missing Baryons particularly challenging.

Direct detection of the emission signal from the WHIM requires ideally large field of view and effective area imager-spectrometers, a combination that is not yet available. The most promising observational strategy with current instrumentation, is to search for the WHIM in absorption (as discrete absorption lines) in the UV and x-ray spectra of bright, intrinsically featureless, background astrophysical sources.

At WHIM temperatures the dominant ions of oxygen (the most abundant metal in gas with solar-like composition) are OVII ($\gtrsim$ 80 % of oxygen in the entire WHIM temperature range) and OVI ($\lesssim$ 20 % of oxygen at the low end of the WHIM temperature distribution). The strongest transitions from these ions are the OVI $1s^2 2s \to 1s^2 2p$ doublet at 1031.9 and 1037.6 Å (UV) and the OVII K$\alpha$ resonant at 21.6 Å (x-rays). The key to the detectability of an absorption line is the transition contrast, i.e. the ratio between the line wavelength and its equivalent width (EW), compared with the spectrometer resolving power. The line transition contrast expected from the WHIM is similar in the UV and the x-ray band: [1031.9 / EW(OVI$_{1031.9}$)] $\gtrsim$ 1500-15000 versus [21.6 / EW(OVII$_{21.6}$)] $\gtrsim$ 2700-27000. However, although the resolving power of the current UV spectrometers is similar to or better than the OVI transition contrast, in the x-ray band the situation is worse. Consequently, despite the dominant distribution of OVII expected in the WHIM (tracking 70-80 % of the WHIM mass) compared to OVI, the search for the WHIM, so far, has been much more fruitful in the UV than in the x-ray band [e.g. (*7,13*)].

Hydrodynamical simulations for the formation of LSS in the universe [e.g. (*8*)], successfully reproduce the measured number of OVI absorbers per unit redshift interval with EW larger than a given threshold (*7,9*) . Danforth & Shull (*7*) estimated the contribution of the WHIM baryons detectable through OVI absorption down to an OVI column density of $N_{OVI} > 10^{13.4}$ cm$^{-2}$ (i.e. the low temperature end of the WHIM mass distribution) to $\Omega$, and found that it is, at most, only 10 % of the missing baryons ($\Omega_{OVI}$ = 0.22 %).

An alternative, and complementary, way of searching for the "missing mass" in the local universe, is to look for hydrogen absorption in the form of broad Ly$\alpha$ absorbers (BLAs). At WHIM temperatures most of the H is ionized, but the residual HI (neutral hydrogen) can still imprint Lyman series absorption onto the UV spectra of background objects; these absorption lines must therefore be broad (with widths given by Doppler parameter $b \geq 42$ km s$^{-1}$, for temperatures $\log T \geq 5$). Richter and collaborators (*13*) analyzed HST spectra from the lines of sight to 4 bright active galactic nuclei (AGN), and found that the fraction of $\Omega$ detectable through BLAs down to $N_{HI}/b > 10^{11.3}$ cm$^{-2}$ km$^{-1}$ s is $\Omega_{BLAs} = 0.27$ %, similar to the contribution to $\Omega$ detectable through OVI absorbers, and again only ~ 10 % of the 'missing baryons'. These studies prove not only that both BLAs and OVI absorbers can be considered potential good tracers of the WHIM, but also that the WHIM itself has mostly likely been already detected through these two transitions in the UV band. However, owing to the high temperature of the WHIM, only a relatively small fraction of the WHIM distribution can be probed by BLAs and OVI absorbers: the vast majority (70 to 80 %) of the missing mass remains to be found via the x-ray band.

Owing to to the observational difficulty of making these measurements, they are controversial. The statistically strongest evidence to date comes from a *Chandra* LETG (low-energy transmission grating spectrometer) spectrum of the nearby (z=0.03) blazar Mkn 421, that shows a number of absorption lines, some identified with two different intervening WHIM systems at *z*=0.011 and *z*=0.027 (*14,15*) at significances of 3.5σ and 4.9σ, respectively. However, analysis of a better signal to noise (S/N), but poorer spectral resolution, XMM-*Newton* reflection grating spectrometer (RGS) spectrum of Mkn 421 does not show some of these absorption lines (*16-18*). We note (*16*) however that both sets of observations are consistent within observational errors (Figure 2) Further observations are thus required to definitely confirm or rule out the proposed WHIM identifications along this line of sight.

Deep WHIM studies require both the UV and the x-ray bands. The x-ray band is crucial to detect the WHIM systems and derive an accurate ionization correction. The UV band is vital to measure the associated amount of HI and hence the baryonic mass in each system, which otherwise depends on the fraction of heavier elements to hydrogen (the "abundances"). The full potentiality of UV WHIM studies will be available only after the Shuttle Servicing Mission 4 has restored spectral capabilities to the HST: the new COS UV spectrograph will provide spectra of accurately selected bright x-ray targets with S/N sufficient to detect BLAs. In the x-ray band, the situation is far more challenging because future instruments are some way off. Current observational facilities should be first exploited to their limits.

According to theory, chances of finding a WHIM filament along a random line of sight in the universe increases with (i) the crossed path length, between us and the beacon used to obtain x-ray images of the intervening space; and (ii) the inverse of the baryon column density in the filament: the larger the amount of baryons in the filament, the lower the probability of finding one. Observationally, instead, the chances of detecting such systems increase with the square root of (i) the detector throughput (the effective area $A_{eff}$), (ii) the spectrometer resolving power ($R = \lambda/\Delta\lambda$), and (iii) the amount of dedicated observing time ($\Delta t$); for a given instrument, $A_{eff}$ and $R$ are fixed parameters, and only the observation exposure $\Delta t$ can be chosen.

It can be shown that detection of the WHIM is within the reach of current instrumentation, but it requires very long observations, on the order several million seconds. Hydrodynamical simulations predict (*9*) a 95.5 % chance of finding at least one OVII WHIM system with column density $N_{OVII} \geq 8 \times 10^{14}$ cm$^{-2}$, along any random line of sight between z=0 to 0.3. This probability raises to 99.7 % when $N_{OVII}$ is less than half this value. For background targets with a quiescent soft x-ray flux of $\geq 1$ mCrab (of which there are few), ~2-3 Ms and ~ 10 Ms are required, with either the LETG or the RGSs, to become sensitive, at a statistical significance $\geq 3\sigma$, to $N_{OVII} = 8 \times 10^{14}$ cm$^{-2}$ and $N_{OVII} = 4 \times 10^{14}$ cm$^{-2}$, respectively. Combined LETG and RGSs exposures of up to ~10 Ms, therefore, would certainly allow us to either detect the WHIM or to disprove models.

Deep WHIM studies, however, await future missions. *Constellation*-X (*19*) [and at a similar extent *XEUS* (*20*), or any of the several proposed mission concepts dedicated to WHIM studies, e.g. *Pharos* (*21*) or *Edge* (*22*)) will allow us to study ~ 60 different lines of sights towards all the available extragalactic targets at z$\geq$0.3 and with quiescent soft x-ray flux $\geq 0.2$ mCrab, with average exposures of $\leq 100$ ks each. Lowering the flux threshold by a factor of only 2 (from 0.2 to 0.1 mCrab) would more than double the number of available targets, and hence lines of sights, requiring only twice as deep exposures per line of sight.

Alternative baryon reservoirs had been proposed before hydro-dynamical simulations, run in the framework of a Λ-CDM universe, became available, which include, for example, large amounts of cold molecular gas around galaxies (*23*) or tenuous intragroup hot gas [e.g. (*7*)]. However, these scenarios clash with the outcome of Λ-CDM simulations, which all agree in predicting the WHIM as the main reservoir of baryons in the local universe. In the next decade we will be in a position to trace all the WHIM baryons in the universe and so either see whether the missing baryon problem still holds or validate theoretical predictions and the SCM.

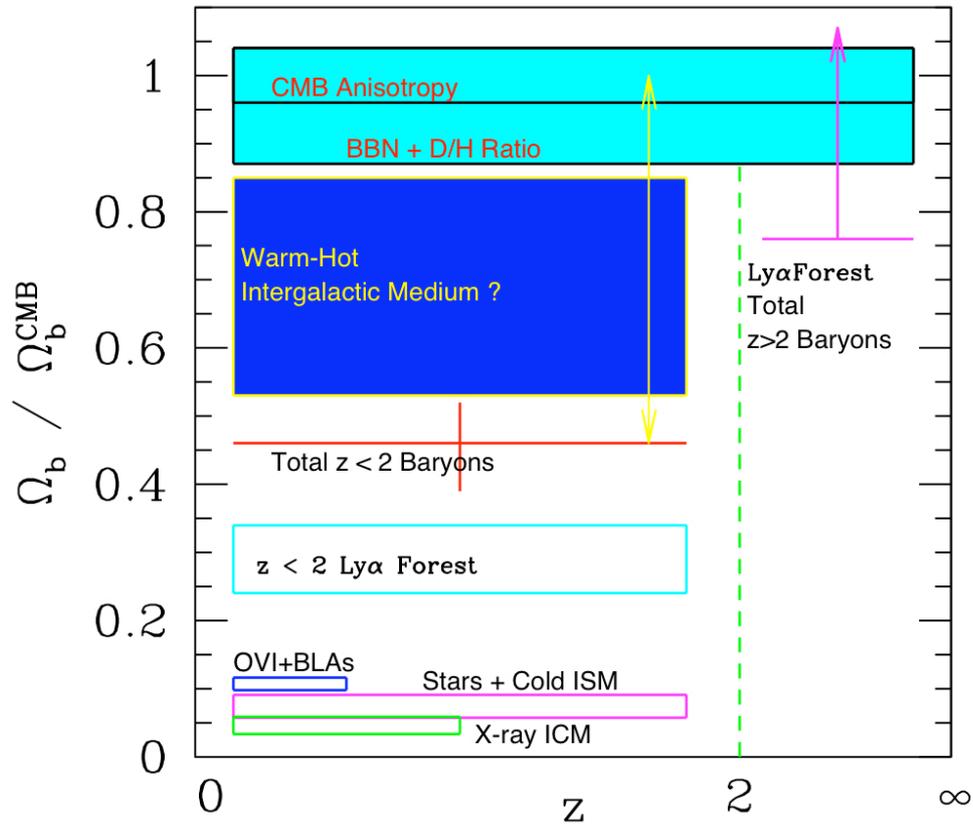

**Figure 1 The "Missing Baryons" and the Warm-Hot Intergalactic Medium**

Baryon density in the universe, at all redshifts, normalized to the cosmological mass density of baryons derived from cosmic microwave background (CMB) anisotropy measurements. Two completely independent measures, based on anisotropies of the CMB on one hand, and on observations of the deuterium-to-hydrogen (D/H) ratio combined with big bang nucleo-synthesis (BBN) models, on the other, nicely converge toward a total cosmological mass density of baryons in the universe, of ~4.5 % (cyan shaded rectangle). At redshifts z > 2 nearly all the expected baryons are found in the Lyα Forest (magenta lower limit: the measure is a lower limit due to uncertainties in the ionization correction). At $z \leq 2$, however, by adding up the baryons observed in stars and cold interstellar medium (ISM) in galaxies, residual low-$z$ Lyα Forest, OVI and BLA absorbers, and x-ray hot gas in clusters of galaxies, account for less than half of the expected cosmological mass density of baryons in the universe (red point). The rest of the baryons still elude detection, and hence are "missing baryons". A theoretical solution to this problem has been relatively recent offered by hydro dynamical simulations for the formation of structures in the universe.

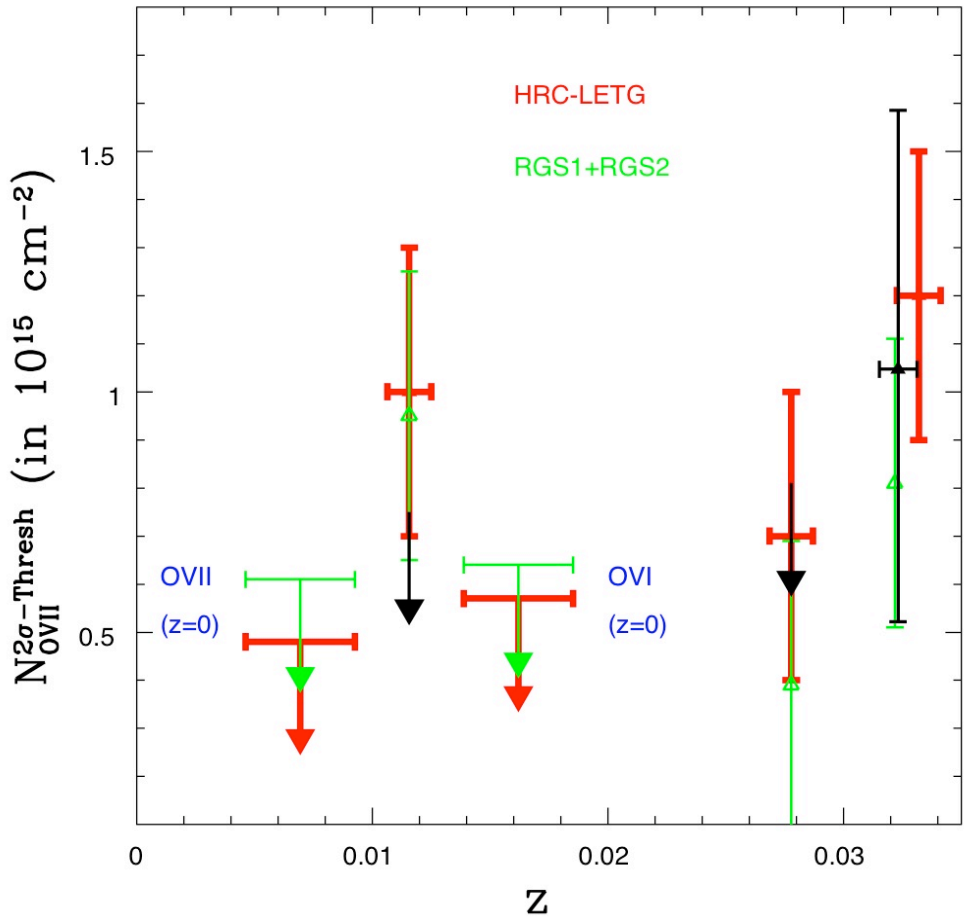

**Figure 2: XMM-*Newton* RGS Spectra of Mkn 421 are consistent with the two *Chandra* LETG Detections of the WHIM.**

Measurements (with associated 1σ errors) and 2σ upper limits of OVII column densities from z=0 to z=0.03 (the redshift of Mkn 421), as derived from the *Chandra* LETG (red symbols) spectrum of Mkn 421 (*14,15*) and our (green symbols) and Rasmussen et al.'s (*17*) (black symbols) analyses of the XMM-*Newton* RGS spectrum of Mkn 421. The two bins at z=0 to 0.004 and z=0.020 to 0.024 contain absorption lines both in the LETG and RGSs spectra, noncontroversially identified as $z=0$ OVII and OVI transitions, and are not shown in this figure. In Rasmussen et al.'s analysis of the RGS spectrum, the upper limits at $z=0.011$ and $z=0.027$ are fully consistent with the Chandra measurements within their $1\sigma$ statistical errors. Moreover, in our analysis of the RGS spectrum of Mkn 421, we actually detect absorption lines at these two positions (green points at $z=0.011$ and $z=0.027$), with intensities consistent with those of the lines detected in the Chandra spectrum of Mkn 421.